\documentstyle[emulate_apj,apjfonts,epsf]{article}
\gdef\h50min{$h_{50}^{-1}$}
\gdef\kms{km\,s$^{-1}$}
\gdef\1054{MS\,1054$-$03}
\gdef\2053{MS\,2053$-$04}
\lefthead{van Dokkum et al.}
\righthead{Evolution of Field Early-Type Galaxies}
\slugcomment{Accepted for publication in the Astrophysical Journal Letters}

\begin{document}

\title{Luminosity Evolution of Field Early-Type Galaxies to $z=0.55$
\altaffilmark{1,2}}
\author{Pieter G. van Dokkum \altaffilmark{3,4}, Marijn Franx \altaffilmark{5},
Daniel D. Kelson \altaffilmark{6},
and Garth D. Illingworth \altaffilmark{7}}
\altaffiltext{1}
{Based on observations with the NASA/ESA {\em Hubble Space
Telescope}, obtained at the Space Telescope Science Institute, which
is operated by AURA, Inc., under NASA contract NAS 5--26555.}
\altaffiltext{2}
{Based on observations obtained at the W.\ M.\ Keck Observatory,
which is operated jointly by the California Institute of
Technology and the University of California.}
\altaffiltext{3}{California Institute of Technology, MS105-24, Pasadena,
CA 91125}
\altaffiltext{4}
{Hubble Fellow}
\altaffiltext{5}{Leiden Observatory, P.O. Box 9513, NL-2300 RA, Leiden, The
Netherlands}
\altaffiltext{6}{OCIW, 813 Santa Barbara St., Pasadena, CA 91101}
\altaffiltext{7}{University of California Observatories/Lick Observatory,
University of California, Santa Cruz, CA 95064}

\begin{abstract}

We study the Fundamental Plane (FP) of field early-type galaxies
at intermediate redshift, using HST WFPC2 observations and deep
Keck spectroscopy. Structural parameters and internal
velocity dispersions are measured for eighteen galaxies
at $0.15<z<0.55$. Rest frame $M/L_B$ ratios are
determined from the Fundamental Plane and compared to those
of cluster early-type galaxies at the same redshifts.
The systematic offset between $M/L$ ratios
of field and cluster early-type galaxies at intermediate redshift
is small, and not significant:
$\langle \ln M/L_B \rangle_{\rm field} - \langle \ln M/L_B \rangle_{\rm
clus} = -0.18 \pm 0.11$.
The $M/L_B$ ratios of field early-type galaxies evolve
as $\Delta \ln M/L_B = (-1.35 \pm 0.35) \times z$, very similar
to cluster early-type galaxies.
After correcting for luminosity evolution,
the FP of field early-type galaxies has a scatter
$\sigma_{\rm BI} = 0.09 \pm 0.02$ in $\log r_e$, similar to
that in local
clusters. The scatter appears to be driven by low mass S0 galaxies;
for the elliptical galaxies alone we find $\sigma = 0.03^{+0.04}_{-0.03}$.
There is a hint that the FP has a different slope than
in clusters, but more data are needed to confirm this.
The similarity of the $M/L$ ratios of cluster and field
early-type galaxies provides a constraint on the relative
ages of their stars. At $\langle z \rangle = 0.43$,
field early-type galaxies are younger than cluster
early-type galaxies by only $21 \pm 13$\,\%, and we infer
that the stars in field early-type galaxies probably formed at
$z \gtrsim 1.5$. Recent
semi-analytical models for galaxy formation in a $\Lambda$-CDM Universe
predict a systematic difference between field
and cluster galaxies of $\Delta \ln M/L_B \sim -0.6$,
much larger than the observed difference.
This result is consistent with the hypothesis that field
early-type galaxies formed earlier than predicted
by these models.

\end{abstract}


\section{Introduction}

\begin{figure*}[t]
\epsfxsize=18.5cm
\epsffile{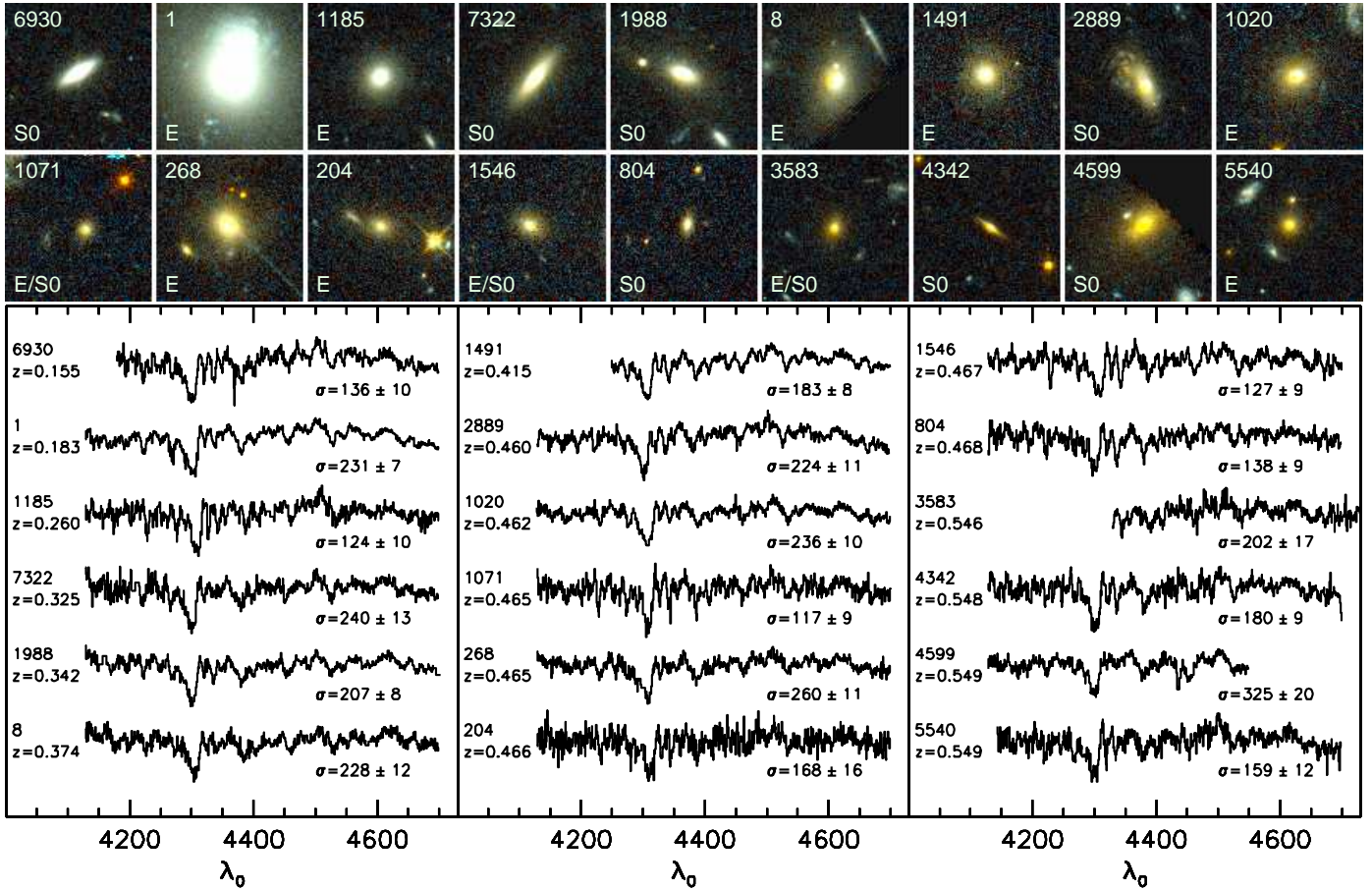}
\caption{\small
HST WFPC2 images and Keck spectra of field early-type galaxies
with measured velocity dispersions, ordered by increasing redshift.
The images are $13\arcsec \times 13\arcsec$. The observed
$R_{F606W}-I_{F814W}$ color of the galaxies becomes
redder at higher redshift, because of the systematic
shift of the rest frame 4000\,\AA{} break.
The spectra were not smoothed or binned. The most prominent feature
in the spectra is the G-band at $4300$\,\AA.
\label{data.plot}}
\end{figure*}

Early-type galaxies in nearby clusters obey
a tight relation between their structural parameters
and central velocity dispersions, known as the ``Fundamental Plane'' (FP)
(Djorgovski \& Davis 1987; Dressler et al.\ 1987).
The redshift evolution of the
zeropoint of the relation is proportional to the evolution of the
mean $M/L$ ratio, and is a very sensitive indicator of the
mean age of the stars in early-type galaxies (van Dokkum \& Franx 1996)
[vDF].
The zeropoint evolves very slowly, indicating that the stars in massive
early-type galaxies formed at $z \geq 3$ (e.g., Kelson et al.\ 1997,
Bender et al.\ 1998, van Dokkum et al.\ 1998).
This result is consistent with other studies on the evolution of
luminosities, colors,
and line strengths of early-type galaxies (e.g., Bender, Ziegler,
\& Bruzual 1996, Ellis et al.\ 1997, Stanford et al.\ 1998, 
Barger et al.\ 1998, Kelson et al.\ 2001).

Such old stellar populations of early-type galaxies seem difficult to
reconcile with hierarchical galaxy formation models, which place the assembly
time of massive early-type galaxies at much lower redshift (e.g.,
Kauffmann 1996, Baugh, Cole, \& Frenk 1996).  However, the assembly
time of early-type galaxies
can be much later than the mean formation time of
their stars.
Furthermore, the true luminosity and color evolution of
early-type galaxies can be underestimated as a result of morphological
evolution (the ``progenitor bias''; van Dokkum \& Franx 2001).  As a
consequence, current semi-analytical hierarchical models are able to
reproduce the slow evolution of the $M/L$ ratios of cluster
early-type galaxies as well as the low scatter in their color-magnitude
relation (e.g., Diaferio et al.\ 2001).

A long standing prediction of hierarchical models is that early-type
galaxies in the {\em field} are younger than those in
rich clusters,
because galaxy formation is accelerated in dense environments (e.g.,
Kauffmann 1996, Baugh, Cole, \& Frenk 1996).
In this {\em Letter} we present the Fundamental Plane
of field early-type galaxies at intermediate redshift, and compare the
$M/L$ ratios and ages
of early-type galaxies in the field to those in rich clusters.
The FP of distant field early-type galaxies has been studied
previously by Treu et al.\ (1999), who observed six galaxies at
$z \approx 0.3$, and by Kochanek et al.\ (2000), who studied
the $M/L$ ratios of gravitational lenses at $0<z<1$.
Our work is complementary to these studies, and 
has the advantage that the methods are identical to those
used for the analysis of the FP in distant clusters. Therefore, the
field and cluster data can be compared directly, with small systematic
uncertainty. We assume $\Omega_m=0.3$, $\Omega_{\Lambda}=0.7$,
and $H_0 = 50$\,\kms\,Mpc$^{-1}$.

\section{Imaging}

Field early-type galaxies were selected from HST WFPC2 mosaics centered
on the
clusters \2053{} ($z=0.58$) and \1054{} ($z=0.83$).  The data were
taken as part of our ongoing program to obtain deep, large format
images of distant X-ray selected clusters.
The HST mosaic of the cluster \1054{} at $z=0.83$ was presented
in van Dokkum et al.\ (2000).  The \2053{} field was observed with HST on
several occasions in September -- December 1998. The mosaic consists
of six slightly overlapping pointings in the $I_{F814W}$ and
$R_{F606W}$ filters.  Exposure times were 3200\,s in $I_{F814W}$ and
3300\,s in $R_{F606W}$ at each position.  The reduction procedures were
identical to those used for the \1054{} mosaic (cf.\ van Dokkum
et al.\ 2000).
Morphological classifications of all galaxies to $I_{F814W} = 22$
in both HST mosaics are given in Fabricant et al., in
preparation. The
classification system is described in Fabricant, Franx, \& van
Dokkum (2000).

Structural parameters were derived from the $I_{F814W}$ images
by fitting $r^{1/4}$ models, convolved with the
WFPC2 PSF, directly to the 2D light distribution, as
described in vDF.
Observed $I_{F814W}$ surface brightnesses were corrected for
galactic extinction and transformed to
rest frame $B$ using the observed $R_{F606W} - I_{F814W}$ colors
(see vDF).

\section{Spectroscopy}

Field early-type galaxies with $I<21.5$ in the \2053{}
field were observed August 2 -- 3 1997, with
LRIS (Oke et al.\ 1995) on the Keck II telescope.
The \1054{} field was observed January 23 and
February 12 1999. In the selection process confirmed cluster
galaxies (van Dokkum et al.\ 2000)
were excluded, and confirmed field galaxies were given
higher priority than galaxies without redshift.
No color selection was applied.
Twenty-three field early-type galaxies were observed.
The 600\,lines\,mm$^{-1}$ grating was used, with slit widths
of $1\farcs 0$.
Exposure times ranged from 2400\,s to
16200\,s, depending on the magnitudes of objects in the masks.
The reduction procedures were very similar to those described
in vDF and Kelson et al.\ (2000).

Velocity dispersions were determined from a fit to
template star spectra in real space, following the procedures
outlined in vDF and Kelson et al.\ (2000).
For each galaxy high resolution
template star spectra were smoothed to match the
(wavelength dependent) instrumental resolution. Typically,
the instrumental resolution $\sigma_*=
60 - 90$\,\kms{} and the fitting region was
4130 -- 4700 \AA{} in the rest frame.
Velocity dispersions were corrected to a $3\farcs 4$ aperture
at the distance of Coma (cf.\ J\o{}rgensen, Franx, \& Kj\ae{}rgaard
1995).
From experimenting with the choice of template star, the fitting
region, and the polynomial filtering of the spectra, we estimate the
velocity dispersions have a systematic uncertainty of $\sim 5$\,\%.
Spectra and WFPC2 images of the eighteen early-type galaxies with reliable
velocity dispersions (i.e., with a random error in $\sigma$ less
than $10$\,\%) are shown in Fig.\ \ref{data.plot}.

\section{Evolution of the Mean $M/L$ Ratio}

In rich clusters all galaxies in the sample are at the same distance
and the evolution of the mean $M/L$ ratio can be determined directly
from the zeropoint of the Fundamental Plane (see, e.g., vDF).
Our field sample spans a large redshift range, and we
take a different approach.
The residual from the FP of nearby clusters is determined
for each galaxy individually, and expressed as an offset
in $M/L$ ratio. 
The local ($z<0.04$) cluster FP has the form
\begin{equation}
r_e \propto \sigma^{1.20} I_e^{-0.80}
\end{equation}
in the $B$ band (J\o{}rgensen, Franx, \& Kj\ae{}rgaard 1996). Assuming
that early-type galaxies are
a homologous family the existence of the FP implies that
$
M/L \propto   \sigma^{0.50} r_e^{0.25} \propto M^{0.25}
$
(Faber et al.\ 1987). For a single galaxy, the
residual from the FP
is therefore related to an offset in $M/L$ ratio through
\begin{equation}
\Delta M/L_B \propto \sigma^{1.50} r_e^{-1.25} I_e^{-1}.
\end{equation}

\vbox{
\begin{center}
\leavevmode
\hbox{%
\epsfxsize=7cm
\epsffile{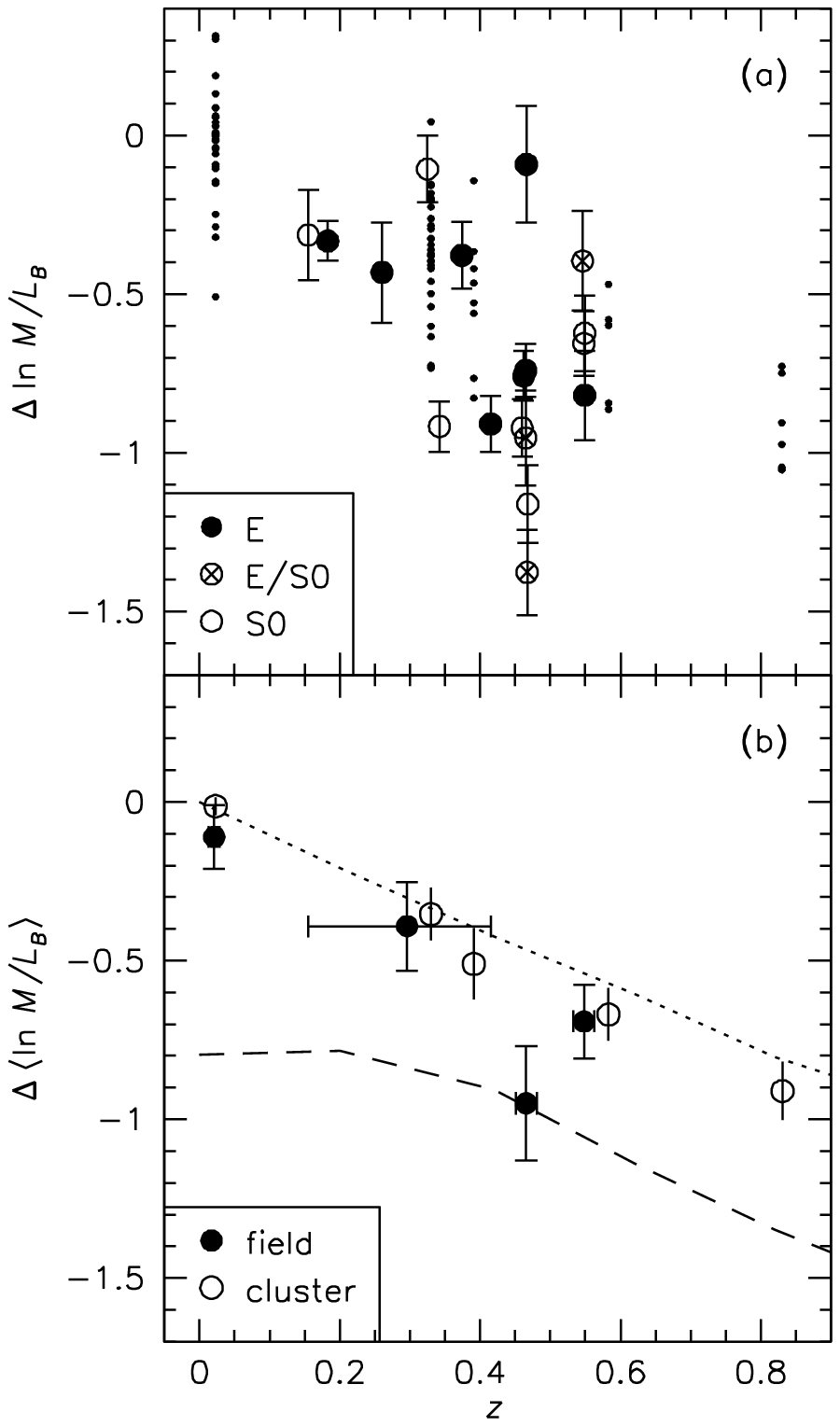}}
\figcaption{\small
(a) Evolution of the $M/L_B$ ratio of field early-type galaxies
(large symbols). Small symbols are
cluster galaxies. (b) The same
data as in (a), binned in redshift. The filled symbol at
$z=0.02$ is derived from data on local galaxies of
Faber et al.\ (1989).  The $M/L$ ratios
of field early-type galaxies are very similar to those of cluster
galaxies. The curves indicate predictions from semi-analytical models
of galaxy formation (Diaferio et al.\ 2001). The dotted line
is for cluster galaxies, and is normalized to Coma. The dashed line is
for field galaxies residing in halos of mass $<10^{14} M_{\odot}$.
\label{evoml.plot}}
\end{center}}

The evolution of the $M/L_B$ ratio of field early-type galaxies is
shown in Fig.\ \ref{evoml.plot}(a) (large symbols).
For comparison, small symbols are
galaxies in Coma at $z=0.02$ (J\o{}rgensen et al.\ 1996)
and in distant clusters (van Dokkum et al.\
1998; Kelson et al.\ 2000). The cluster data were
transformed to the rest frame $B$ band in the same way
as the field galaxies.
In Fig.\ \ref{evoml.plot}(b) the data are binned in redshift.
We used the biweight estimator (Beers, Flynn, \& Gebhardt 1990)
to determine the center of the distribution of
$\Delta \ln (M/L_B)$ and its associated error in each redshift bin.
The filled symbol at $z=0.02$ is derived from the Faber et al.\
(1989) data, by considering galaxies at $z>0.015$ and
assigning galaxies in groups with dispersions $<450$\,\kms{}
to the field. This division corresponds
to a halo mass of $\sim 10^{14} M_{\odot}$.

The $M/L$ evolution is very similar to that of cluster early-type
galaxies. We find $\Delta \langle \ln M/L_B \rangle = (-1.35 \pm
0.35) \times z$, compared to $\Delta \langle \ln M/L_B \rangle =
(-1.12 \pm 0.11) \times z$ for clusters (van Dokkum et al.\ 1998).  We
determine the mean systematic offset
between cluster and field early-type galaxies
by subtracting the best fitting relation for the cluster early-type
galaxies from our field data at $0.15<z<0.55$.
The resulting offset
$\Delta \langle \ln M/L_B \rangle_{\rm field} - \Delta \langle \ln
M/L_B \rangle_{\rm clus} = -0.18 \pm 0.11$.
The offset is small, and not significant.
The uncertainty was determined from Monte-Carlo
simulations, taking both random and systematic errors into account.
Random errors dominate the error budget, because most of the
systematic errors cancel.
We verified that the relative offset between the cluster and field does not
change significantly if the shape of the FP is calculated by minimizing
in either $\log r_e$, $\log \sigma$, or $\log I_e$.
This indicates that selection biases
are not significant.
Both the evolution and the offset indicate that
field galaxies evolved in almost the same way as 
cluster galaxies, and had a comparable formation history.

\section{Slope and Scatter of The Fundamental Plane}

Figure \ref{fp.plot} shows the FP of field early-type
galaxies, with surface brightnesses corrected for luminosity evolution
as determined from the cluster data.  The line is the FP of ten nearby
clusters (J\o{}rgensen et al.\ 1996),
and small dots are galaxies in distant clusters.
Note that most galaxies in the distant cluster sample are in
a single cluster (CL\,1358+62 at $z=0.33$) which has a slightly higher
mean $M/L$ ratio than the other clusters.

The residual offset of the field galaxies
is $0.065$ in $\log r_e$, and is the
result of the small systematic difference in $M/L$ ratio between field
and cluster galaxies. The
scatter in $\log r_e$ is $0.09 \pm 0.02$
after correcting for measurement errors,
similar to the scatter of $0.073$ measured for nearby clusters
(J\o{}rgensen et al.\ 1996).
The scatter for the field elliptical galaxies alone is
only $0.03^{+0.04}_{-0.03}$.

There is a hint that the FP of field early-type galaxies has
a different form than that in (nearby and distant) clusters.
It is hazardous to determine the coefficients of the FP from
our sample, because of selection effects and systematic errors.
Nevertheless, the algorithm of J\o{}rgensen et al.\
(1996) gives
\begin{equation}
r_e \propto \sigma^{1.10 \pm 0.14} I_e^{-0.59 \pm 0.09}
\end{equation}
with scatter $0.08$ in $\log r_e$.
This form of the FP may indicate that the $M/L$ ratios of
field early-type galaxies scale with their structural
parameters as $M/L \propto M^{0.07}r_e^{0.62}$.
An alternative explanation
is that field S0 galaxies evolve
differently from field elliptical galaxies: for elliptical
galaxies alone, there is no indication of a steeper slope.  A third
possibility is that the redshift evolution is inhomogeneous.  Eleven
of the eighteen
field early-type galaxies are in two sheets or filaments with
very small velocity dispersion ($\sigma \lesssim 200$\,\kms)
at $z=0.46$ and $z=0.55$, and there is marginal evidence in Fig.\
\ref{evoml.plot} that there is a systematic offset in $M/L$ ratio
between the galaxies in these structures. The present data are
insufficient to distinguish between these possibilities.

\vbox{
\begin{center}
\leavevmode
\hbox{%
\epsfxsize=7cm
\epsffile{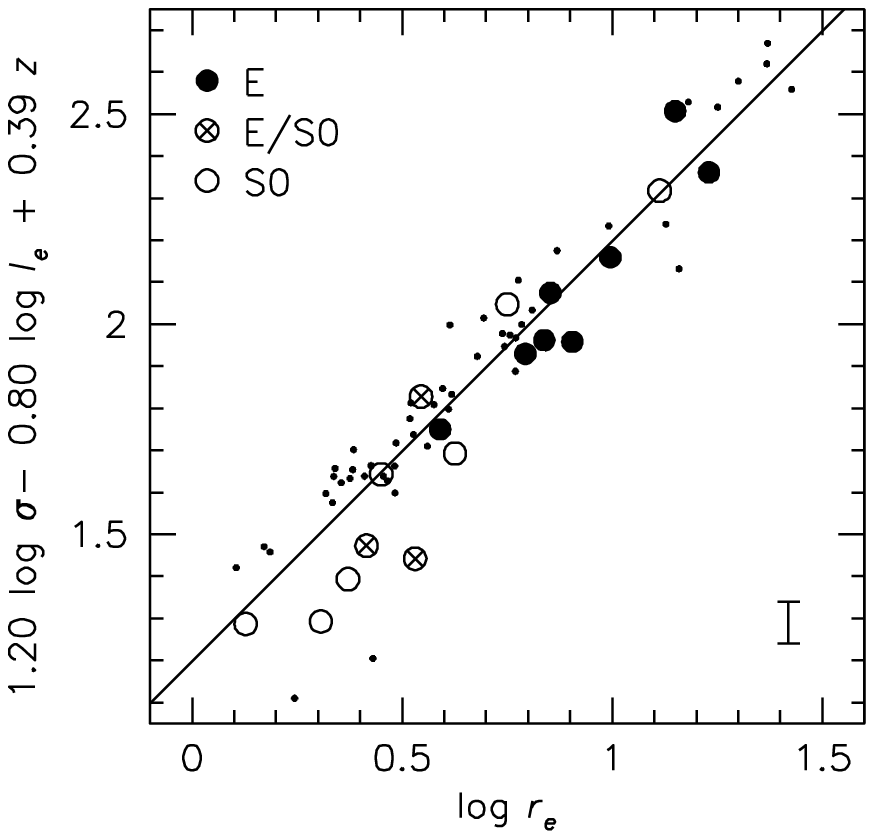}}
\figcaption{\small
The Fundamental Plane at intermediate redshift,
with surface brightnesses of galaxies corrected
for luminosity evolution as determined from the cluster galaxies.
Large symbols are field galaxies; small symbols are
distant cluster galaxies from van Dokkum et al.\ (1998)
and Kelson et al.\ (2000). The line is the FP of nearby clusters,
from J\o{}rgensen et al.\ (1996).
Field galaxies have a small residual
offset from the FP of cluster galaxies. The scatter is
mainly driven by the S0 galaxies. There is a hint that the
form of the FP is different from that in clusters.
\label{fp.plot}}
\end{center}}

\section{Discussion}

The small systematic
offset of $\langle \ln M/L_B \rangle_{\rm field} - \langle \ln
M/L_B \rangle_{\rm clus} = -0.18 \pm 0.11$
corresponds to
an age difference at $\langle z \rangle =0.43$
of $21 \pm 13$\,\%
for solar metallicity and a Salpeter (1955) IMF, with field galaxies
being younger than cluster galaxies
(Worthey 1994; Vazdekis et al.\ 1996). Since the mean
luminosity-weighted formation redshift of the stars in cluster
early-type galaxies $\langle z_{*,\rm clus} \rangle \gtrsim 2$
for $\Omega_m=0.3$, $\Omega_{\Lambda} =0.7$ (van Dokkum \& Franx
2001), the stars in field early-type galaxies probably formed at
$\langle z_{*,\rm field} \rangle \gtrsim 1.5$. 

Our results are consistent with those of Burstein, Faber, \& Dressler
(1990) and Pahre, de Carvalho, \& Djorgovski (1998) for nearby
galaxies, and those of Treu et al.\ (1998) and Kochanek et al.\ (2000)
for distant galaxies. However,
the offset between cluster and field galaxies is
somewhat smaller than that found by de Carvalho \& Djorgovski (1992)
for nearby galaxies. 
Our FP results are consistent with
studies of other age estimators at low redshift (e.g., Bernardi et
al.\ 1999) and high redshift (e.g., Schade et al.\ 1996, 1999, Im et
al.\ 2001).  The FP has the advantages that it has a low
scatter, and that
it measures $M/L$ ratios rather than total luminosities.

It is interesting to compare the observed evolution of the mean $M/L$
ratio to predictions of hierarchical galaxy formation models.  In
Fig.\ \ref{evoml.plot} model predictions of Diaferio et al.\ (2001)
are plotted for cluster and field data. The predictions were generated
from the publically available simulation output
catalogs\footnote{www.mpa-garching.mpg.de/Virgo/data\_download.html},
by selecting galaxies with
$M_B<-21.5$ and bulge-to-total light ratio $B/T>0.4$, and using a
boundary halo mass of $10^{14} M_{\odot}$ to distinguish cluster
galaxies and field galaxies (cf.\ Diaferio et al.\ 2001).  The models
predict an offset of $\Delta \ln M/L_B \sim -0.6$ between field and
cluster early-type galaxies, almost independent of redshift. The
offset remains unchanged when only galaxies with $B/T>0.8$
are considered.  The
predicted difference is significantly larger than the observed
difference, at all redshifts. This may indicate that
field early-type galaxies were generally formed
at much higher redshift than predicted by these models,
or that late star formation is prevented by some
mechanism which is not included in the models.

In our estimates of the relative ages of field and cluster galaxies
we have assumed that they have the same metallicity.
This assumption can be tested by observations
of absorption line strengths of high redshift
field and cluster galaxies.
Furthermore, we have implicitly assumed
that the high redshift data and the low redshift data can be compared
directly, which may not be the case if field early-type galaxies
experienced significant morphological evolution at low redshift
(van Dokkum \& Franx 2001). Our measurement of
the age difference between cluster and field galaxies is
fairly insensitive to these effects, but the
rate of evolution of the $M/L$ ratio may be underestimated.
Nevertheless, the
observed evolution of the $M/L$ ratio can be used
directly to convert luminosities of distant field early-type galaxies
to masses. The combination of the evolution of the $M/L$ ratio and the
luminosity function of field ellipticals (e.g., Schade et al.\ 1999,
Im et al.\ 2001) can give a direct measurement of
the evolution of the mass function of elliptical galaxies, and hence
of their merger history.

\begin{acknowledgements}
We thank Guinevere Kauffmann for her help in obtaining the model
predictions, and the referee, Ralf Bender, for his comments.
P. G. v. D. acknowledges support by NASA through Hubble Fellowship
grant HF-01126.01-99A
awarded by the Space Telescope Science Institute, which is
operated by the Association of Universities for Research in
Astronomy, Inc., for NASA under contract NAS 5-26555.
Support from STScI grants GO06745-95A and
GO07372-96A is gratefully acknowledged.
\end{acknowledgements}

\end{document}